\documentclass[a4paper,conference]{IEEEtran}
 
\IEEEsettopmargin{t}{30mm} 
\IEEEquantizetextheight{c}  
\IEEEsettextwidth{14mm}{14mm}  
\IEEEsetsidemargin{c}{0mm}

\usepackage[utf8]{inputenc}
\usepackage[T1]{fontenc}
\usepackage{url}
\usepackage{tikz}
\usepackage{pgfplots}
\usepgfplotslibrary{groupplots}
\usepgfplotslibrary{dateplot}
\pgfplotsset{compat=newest}
\usepackage{pgf}
\usepackage{ifthen}
\usepackage{cite}
\usepackage[cmex10]{amsmath} 
\usepackage{amsfonts, amssymb}
\usepackage[version=4]{mhchem}
\usepackage{graphicx, amsthm, float, dsfont,color}

\IEEEoverridecommandlockouts

\newtheorem{theorem}{Theorem}
\newtheorem{lemma}{Lemma}

\makeatletter
\newcommand{\vast}{\bBigg@{3}}
\newcommand{\Vast}{\bBigg@{4}}
\makeatother

\newcommand{\bs}[1]{\boldsymbol{#1}}
\newcommand{\mcl}[1]{\mathcal{#1}}

\newcommand{\msf}[1]{\mathsf{#1}}

\newcommand{\Q}{\mcl{Q}}

\makeatletter
\g@addto@macro\@floatboxreset{\setlength{\@fptop}{0pt}\setlength{\@fpsep}{0pt}\setlength{\@fpbot}{0pt}}
\makeatother
\begin{document}
\title{Two-Class Joint Source-Channel Coding: Expurgated Exponents with i.i.d. Distributions}


\author{%
  \IEEEauthorblockN{Seyed AmirPouya Moeini}
  \IEEEauthorblockA{University of Cambridge\\
    \texttt{sam297@cam.ac.uk}}
  \and
  \IEEEauthorblockN{Albert Guill\'en i F\`abregas}
  \IEEEauthorblockA{University of Cambridge\\
  Universitat Polit\`ecnica de Catalunya\\
    \texttt{guillen@ieee.org}}
  \thanks{This work was supported in part by the European Research Council under Grant 101142747, and in part by the Spanish Government under Grant PID2020-116683GB-C22}
}

\maketitle

\begin{abstract}
This paper studies expurgated exponents for joint source-channel coding of discrete memoryless sources and channels under i.i.d. random coding. 
We show that a two-class partitioning of source sequences, 
where the codeword distribution depends on the source type,
achieves an exponent at least as high as that of optimal single-class coding, 
in which the codeword distribution is independent of the source message.
\end{abstract}

\section{Introduction}
We study the transmission of non-equiprobable messages from a discrete memoryless source (DMS) over a discrete memoryless channel (DMC).
In 1980, Csiszár \cite{csiszar1980} showed that joint source-channel coding can achieve a lower error probability than separate source and channel coding, and derived the corresponding random-coding error exponent using a construction in which codewords are drawn from a distribution that depends on the source message.
In contrast, Gallager \cite{gallager} had earlier derived a different error exponent  
 where codewords are drawn from a fixed distribution, independent of the source message.
 Zhong et al. \cite{1614076} later revisited these results and showed that Csiszár's exponent is always at least as large as Gallager’s, and sometimes strictly larger.

Compared to random coding, expurgated bounds in joint source-channel coding have received relatively little attention. 
In the channel coding, the main techniques for obtaining expurgated exponents are due to Gallager \cite[Section~5.7]{gallager} and Csiszár-Körner-Marton (CKM) \cite{1056281, csiszar1977new}.
See \cite{6810903} for a more extensive study of expurgated bounds in channel coding.

In \cite{1056585}, Csiszár generalized the CKM method and derived two expurgated exponents for joint source-channel coding. 
The first corresponds to single-class coding, in which all codewords are drawn from a fixed distribution (a Gallager-like setting), 
while the second assigns a distinct codeword distribution to each source-type class, so that the codeword distribution of a source sequence depends on its type, following \cite{csiszar1980}.
The comparison between these two exponents has remained an open problem.
Scarlett et al. \cite{8006660} later revisited the expurgation problem in joint source-channel coding and rederived Csiszár's exponents using a new technique
based on source-type duplication and type-by-type expurgation.
More recently, Moeini et al. \cite{moeini2025-2} extended this line of work by considering a broader class of message partitionings,
and showed the existence of a two-class partitioning that recovers Csiszár’s second exponent in \cite{1056585}.

To the best of our knowledge, the partitioned and non-partitioned approaches have not been directly compared in the expurgated setting, either theoretically or numerically. 
It has therefore remained open whether partitioning can yield a larger exponent, as it does in the random coding case.

In this work, we focus on i.i.d. random coding ensembles for convenience, in contrast to earlier studies \cite{1056585, csiszar1980, moeini2025-2, 8006660} that considered constant-composition coding. 
We establish the existence of a two-class partitioning that achieves an exponent no smaller than that of optimal single-class coding, thereby settling one side of the comparison. 
In our analysis, single-class coding appears naturally as a special case of partitioning, an observation aligns with intuition but not reflected in previous studies.
The detailed proofs are presented in Section~\ref{sec-proofs}.


\section{Preliminaries}

We study the transmission of non-equiprobable messages from a DMS with distribution $P^k(\bs{v}) = \prod_{i=1}^k P_V(v_i)$, where $\bs{v} = (v_1, ..., v_k) \in \mcl{V}^k$ 
is the source message, and $\mcl{V}$ is a finite discrete alphabet. 
The channel is a DMC given by $W^n(\bs{y}|\bs{x}) = \prod_{i=1}^n W(y_i | x_i)$, $\bs{x}=(x_1, \ldots, x_n) \in \mcl{X}^n$ and $\bs{y}=(y_1, \ldots, y_n) \in \mcl{Y}^n$,
where $\mcl{X}$ and $\mcl{Y}$ are discrete
alphabets with cardinalities $|\mcl{X}|$ and $|\mcl{Y}|$, respectively.
An encoder maps the length-$k$ source message $\bs{v}$ to a length-$n$ codeword $\bs{x}_{\bs{v}}$, which is then transmitted over the channel,
and  decoded as $\hat{\bs{v}}$ at the receiver upon observing the channel output $\bs{y}$.
We refer to $t \triangleq k / n$ as the transmission rate. 
We study the average error probability, defined as
\begin{align}
	p_e \triangleq \mathbb{P}\big[ \bs{V} \neq \hat{\bs{V}}\big].
\end{align}


We say that an error exponent $E>0$ is \emph{achievable} (for any fixed $t$) if there exists a sequence of codes of length $n$ such that the error probability satisfies
\begin{align}
	p_e \leq e^{-n E+o(n)},
\end{align}
where $\lim _{n \rightarrow \infty} o(n) / n=0$.

In this paper, scalar random variables are denoted by uppercase letters, their realizations by lowercase letters, and their alphabets by calligraphic letters. Random vectors are written in boldface.
For two positive sequences $\{f_n\}$ and $\{g_n\}$, we write $f_n \doteq g_n$ if $\lim _{n \rightarrow \infty} \frac{1}{n} \log \frac{f_n}{g_n}=0$, 
and we write $f_n \, \dot{\leq}\,\, g_n$ if $\lim \sup _{n \rightarrow \infty} \frac{1}{n} \log \frac{f_n}{g_n} \leq 0$.

The type of a sequence $\bs{x}=\left(x_1, ..., x_n\right) \in \mcl{X}^n$ is its empirical distribution, defined by
$\hat{P}_{\bs{x}}(x) \triangleq \frac{1}{n} \sum_{i=1}^n \mathds{1}\left\{{x}_i=x\right\} .$
The set of all probability distributions on an alphabet $\mcl{X}$ is denoted by $\mcl{P}(\mcl{X})$, while $\mcl{P}_n(\mcl{X})$ represents the set of empirical distributions for vectors in $\mcl{X}^n$.
For $P_X \in \mcl{P}_n(\mcl{X})$, the type class $\mcl{T}^n(P_X)$ consists of all sequences in $\mcl{X}^n$ with type $P_X$.
We denote by $N_k$ the number of types in $\mcl{V}^k$, which grows polynomially with $k$~\cite[Lemma~2.2]{csiszarkorner}.

Continuing the line of \cite{moeini2025-2, moeini2025-1}, we consider partitioning the source sequences into $m$ disjoint sets, each consisting of one or more full source-type classes.
Each class $c$ is assigned a codeword distribution $Q_c \in \mathcal{P}(\mathcal{X})$, from which the codewords for source sequences in that class are generated i.i.d. according to $Q_c$.
This setting includes Csiszár’s partitioning \cite{csiszar1980, 1056585}, in which each class corresponds to a single source-type class (i.e., $m = N_k$).
Throughout, we use the indices $i$ and $j$ to refer to source types, and $c$ to indicate the class to which a source message (or a source type) belongs under a given partitioning.
Moreover, for a source type $P_i$, we define $R_i \triangleq tH(P_i)$.


\section{Achievable Expurgated Error Exponents}

In this section, we present our main results. We first derive an expurgated exponent for an arbitrary partitioning, and then consider a special case of partitioning into only two classes.

\begin{lemma}\label{lem-main}
	Let $\{\mcl{A}_1, \ldots, \mcl{A}_m\}$ be a given partition of the source sequences, with each class assigned an i.i.d. codeword distribution from $\mcl{Q}_m=\{Q_1,\ldots,Q_m\}$.
	Then, there exists a sequence of codebooks such that, for all choices of $\{\rho_c\}_{c=1}^{m}$ with $\rho_c \geq 1$,
	\begin{align}
		p_e \,\, \dot{\leq} \, \sum_{c=1}^m \, e^{-n E_{\mathrm{x}}^{\prime}\big(Q_c, \Q_m, \rho_c\big) +  E_s^{(c)}(\rho_c, P^k)},
	\end{align}
	where
	\begin{align}\label{eq-lem-main-1}
		E_{\mathrm{x}}'\big(Q, \Q, \rho\big) \! \triangleq  \min_{\tilde{Q}\in\mcl{Q}} 
		 -\rho \log \sum_{(x,\bar{x})} Q(x)\tilde{Q}(\bar{x}) \, e^{-\frac{d_B(x,\bar{x})}{\rho}},
	\end{align}
	and
	\begin{align}
		E_s^{(c)}\big(\rho, P^k\big) \triangleq \log \left(\sum_{\boldsymbol{v} \in \mathcal{A}_c} P^k(\boldsymbol{v})^{\frac{1}{1+\rho}}\right)^{1+\rho},
	\end{align}
	with $d_B(x,\bar{x}) \triangleq -\log \sum_y \sqrt{W(y|x)W(y|\bar{x})}$ denoting the Bhattacharyya distance.
\end{lemma}
A constant-composition counterpart of this result is given in \cite[Corollary~1]{moeini2025-2}. 
The improvement in Lemma \ref{lem-main}, however, is that the minimization in \eqref{eq-lem-main-1} is restricted to the given set of codeword distributions, 
whereas in \cite[Corollary~1]{moeini2025-2} (equivalently \cite{1056585})  it is taken over all possible distributions.

\begin{theorem}\label{thm-main-1}
	For a pair of i.i.d. distributions $\Q=\left\{Q, Q^{\prime}\right\}$, there exists 
	a partition of the source message set into two classes such that
	the following exponent is achievable
	\begin{align}
		E_{J, \mathrm{ex}}^{\mathrm{mc}}(\mcl{Q}, t)
		&\triangleq \sup_{\rho \geq 1} \bigg\{\overline{E}_{\mathrm{x}}'\big(\Q, \rho\big) - tE_s\big(\rho, P_V\big) \bigg\},
	\end{align}
	where $\overline{E}_{\mathrm{x}}'\big(\Q, \rho\big)$ denotes the concave hull of ${E}_{\mathrm{x}}'(\Q, \rho)$ over $\rho \in [1,\infty)$, 
	with ${E}_{\mathrm{x}}'\big(\Q, \rho\big) \triangleq \max_{Q \in \mcl{Q}} \, E_{\mathrm{x}}'\big(Q, \Q, \rho\big)$, and
	$E_s(\rho, P_V) \triangleq \log \left(\sum_{v \in \mcl{V}} P_V(v)^{\frac{1}{1+\rho}}\right)^{1+\rho}$.
\end{theorem}
The partitioning that achieves this exponent is the same as in \cite[Theorem~2]{6803047}, and is given by
\begin{align}
	&\mcl{A}_1 = \Big\{ \bs{v}\in \mcl{V}^k: P^k(\bs{v}) \leq \gamma^k \Big\} \label{eq-part-1}\\
	&\mcl{A}_2 = \Big\{ \bs{v}\in \mcl{V}^k: P^k(\bs{v}) > \gamma^k \Big\}\label{eq-part-2},
\end{align}
for some $ \gamma \in [0,1]$.
Optimizing over the pair of distributions, the optimal expurgated exponent under this partitioning is 
\begin{align}
	E_{J, \mathrm{ex}}^{\mathrm{mc}}(t) \triangleq \max_{\{Q,Q'\}} E_{J, \mathrm{ex}}^{\mathrm{mc}}(\{Q,Q'\}, t). 
\end{align}


\section{Comparison with Single-Class Coding}
We first present the achievable exponent under single-class coding, closely following \cite[Corollary~4]{moeini2025-2}.
\begin{theorem}\label{thm-single-ach}
	There exist single-class codes with i.i.d. ensembles that achieve the following exponent 
	\begin{align}
		E_{J, \mathrm{ex}}^{\mathrm{sc}}(t) \triangleq \sup _{\rho \geq 1}\bigg\{E_{\mathrm{x}}(\rho)-t E_{{s}}\big(\rho, P_V\big)\bigg\},
	\end{align}
	where $E_{\mathrm{x}}(\rho) \triangleq \max_Q E_{\mathrm{x}}(Q,\rho)$, and
	\begin{align}\label{eq-eq-exiid}
		E_{\mathrm{x}}(Q, \rho) =  \, -\rho  \log \sum_{(x,\bar{x})} Q(x) Q(\bar{x})\, e^{-\frac{d_B(x,\bar{x})}{\rho}}.
	\end{align}
\end{theorem}
We use the notation $E_{J, \mathrm{ex}}^{\mathrm{sc}}(Q,t)$ to denote the exponent corresponding to a fixed codeword distribution $Q$.
A key observation is that this result can be viewed as a special case of the two-class partitioning when $Q = Q' = Q^*$, with $Q^*$ denoting the optimal distribution in Theorem~\ref{thm-single-ach}.
This implicitly implies that $E_{J, \mathrm{ex}}^{\mathrm{mc}}(t)  \geq E_{J, \mathrm{ex}}^{\mathrm{sc}}(t)$, as stated in the following theorem.
\begin{theorem}\label{thm-main}
	The optimal expurgated exponent with two-class partitioning is no smaller than that of optimal single-class coding, namely
	\begin{align}
		E_{J, \mathrm{ex}}^{\mathrm{mc}}(t)  \geq E_{J, \mathrm{ex}}^{\mathrm{sc}}(t).
	\end{align}
\end{theorem}
When the optimal distribution is chosen, the i.i.d. and constant-composition channel-coding expurgated exponents coincide, as shown in \cite[Prob.~10.24]{csiszarkorner}. 
Hence, $E_{J, \mathrm{ex}}^{\mathrm{sc}}(t)$ also equals the optimal single-class expurgated exponent under constant composition.

We therefore establish the existence of a two-class partitioning that, under i.i.d. coding, achieves an exponent no smaller than that of optimal single-class coding with constant composition. 
This gives a one-sided comparison between partitioning and non-partitioning in expurgated joint source-channel coding, a result that, to the best of our knowledge, has not appeared previously.


\subsection{Numerical Examples}

In this section, we provide numerical examples of expurgated error exponents in joint source-channel coding, illustrating in particular the impact of partitioning.
We consider a source-channel pair consisting of a binary memoryless source, and a non-symmetric memoryless channel of the form
\begin{align}
	W=\begin{pmatrix}
		1-2\varepsilon & \varepsilon & \varepsilon\\
		\varepsilon & 1-2\varepsilon & \varepsilon\\
		\delta & \delta & 1-2\delta
	\end{pmatrix},
\end{align}
with the parameters $\varepsilon = 0.0001$ and $\delta = 0.1$.
%

For each case, we are given two codeword distributions $\{Q_1,Q_2\}$, and we observe how the two-class partitioning defined in \eqref{eq-part-1}-\eqref{eq-part-2}
for these distributions compares with single-class coding using one of them.
In each figure, we plot the functions $\overline{E}_{\mathrm{x}}'\big(\{Q_1,Q_2\}, \rho\big) - tE_s\big(\rho, P_V\big)$
and $E_{\mathrm{x}}(Q_c,\rho)-t E_{{s}}\left(\rho, P_V\right)$ for $c=1,2$.
For reference, the values $E_{J, \mathrm{ex}}^{\mathrm{mc}}(\{Q_1,Q_2\}, t)$ and $\max_{c=1,2} E_{J, \mathrm{ex}}^{\mathrm{sc}}(Q_c, t)$ are shown as horizontal solid lines,
while the JSCC random-coding exponent \cite[Eq.~(20)]{6803047} is plotted with dot-dashed lines. 
In all three cases, the expurgated exponents exceed the random-coding exponent, although this need not hold in general. 
Depending on the setup, the random-coding exponent can be strictly larger and even tight, coinciding with the sphere-packing bound, as discussed in \cite{csiszar1980, 1614076}.


The first figure illustrates a case with non-optimal codeword distributions, where the partitioned exponent is strictly higher than the two single-class exponents.
\begin{figure}[H]
    \centering
    \makebox[\linewidth][c]{%
        \resizebox{1.1\linewidth}{!}{\input{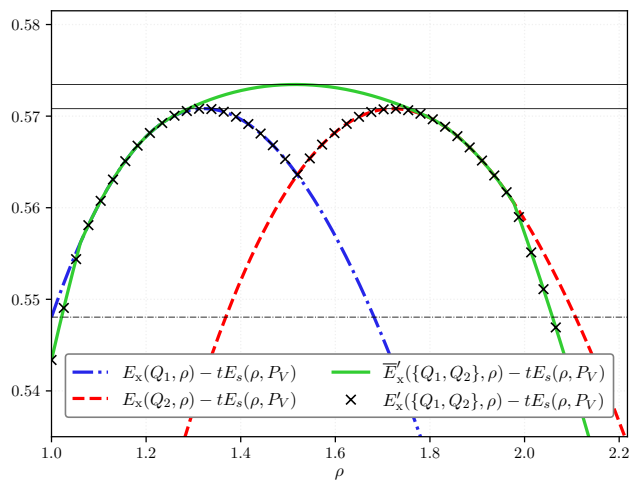}}%
    }
    \vspace{-1.75\baselineskip}
    \caption{Example where partitioning strictly improves the exponent, obtained with $Q_1 = (0.4,0.4,0.2)$, and $Q_2 = (0.5,0.5,0)$, $t = 0.75$ and $P_V(0) = 0.025$.}
        \label{fig-part}
\end{figure}

The second plot, on the other hand, illustrates a case where partitioning weakens the achievable exponent, with the codeword distributions still being non-optimal.

\begin{figure}[H]
    \centering
    \makebox[\linewidth][c]{%
        \resizebox{1.1\linewidth}{!}{\input{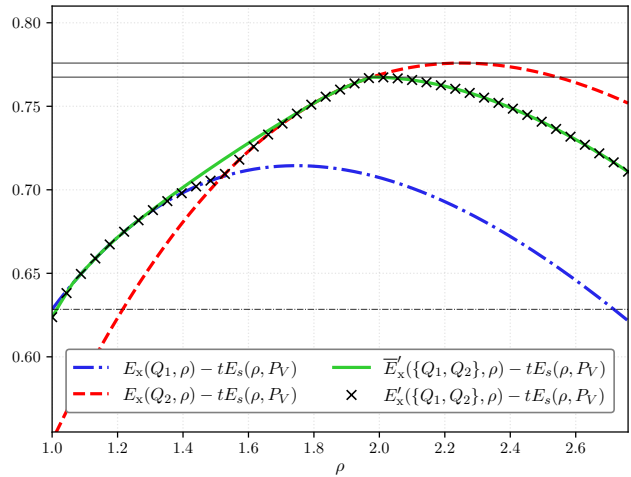}}%
    }
    \vspace{-1.75\baselineskip}
    \caption{Example where single-class coding yields a higher exponent, with the same setting as Fig.~\ref{fig-part} except $t = 0.5$ and $P_V(0) = 0.02$.}
    \label{fig-sng}
\end{figure}

Figures \ref{fig-part} and \ref{fig-sng} show that when the codeword distributions are fixed and not optimized, it is not predetermined which exponent dominates; either one can exceed the other.
In the final example, $Q_1$ and $Q_2$ are chosen optimally, showing that in this particular case the two exponents coincide.


\begin{figure}[H]
    \centering
    \makebox[\linewidth][c]{%
        \resizebox{1.1\linewidth}{!}{\input{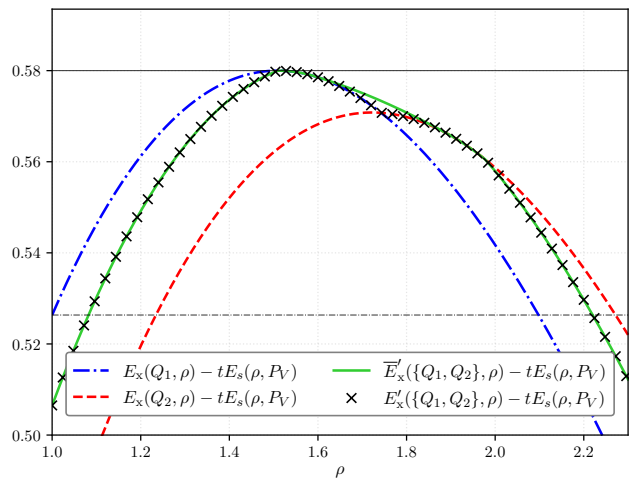}}}
    \vspace{-1.75\baselineskip}
    \caption{The two exponents coincide under optimal distributions, with the same setting as Fig.~\ref{fig-part} except $Q_1 = (0.4489,0.4489,0.1021)$ and $Q_2 = (0.5,0.5,0)$.}
\end{figure}

To date, we have not found any joint source-channel coding setting in which optimal partitioning yields a strictly higher exponent. 
In cases examined, the two exponents coincide, raising the possibility that they may always be equal.
Observe that for a gap to exist, the input alphabet of the channel must have size at least three. 
This is because when the input alphabet size is two, it has been shown 
\cite{1054148}
 that the uniform distribution $Q(0)=Q(1)=1/2$ always maximizes $E_{\mathrm{x}}(Q,\rho)$, regardless of the value of $\rho$. 
Consequently, for binary-input channels, partitioning does not provide any improvement to the exponent.

\section{Proofs} \label{sec-proofs}
The work of \cite{8006660} revisits expurgation in joint source-channel coding and rederives Csiszár's two exponents \cite{1056585} using a new technique based on source-type duplication and type-by-type expurgation. 
The key result of \cite{8006660}, which forms the basis for all achievability results in this paper, is summarized in the following lemma.
\begin{lemma}\label{lem-expurgation-jscc}
	There exists a joint source-channel code $\mcl{C}$ such that for any source type $P_i \in \mcl{P}_k(\mcl{V})$ and any sequence $\bs{v} \in \mcl{T}^k(P_i)$, the source message error probability satisfies
	\begin{align} \label{eq-lem-exp}
		p_e(\bs{v}, {\mcl{C}})  \,\, \dot{\leq} \,\, 
		 \sum_{j=1}^{N_k} \, \mathbb{E}\Big[p_e\big(\bs{v}, j, \mathsf{C}\big)^{\frac{1}{\rho_{ij}}} \Big]^{\rho_{ij}},
	\end{align}
	where $\rho_{ij} > 0$, and $p_e\big(\bs{v}, j, \mathsf{C}\big)$ refers to the probability that, given $\bs{V} = \bs{v}$, there exists some $\bs{\bar{v}} \in \mcl{T}^k(P_j)$
	that yields a decoding metric at least as high as that of $\bs{v}$.
	Moreover, the expectation is taken with respect to the given (not necessarily distinct) codeword distributions $\{P_{\bs{X}}^{(j)}\}_{j=1}^{N_k}$.
\end{lemma}

\subsection{Proof of Lemma \ref{lem-main}}
The proof closely follows the approach in \cite[Lemma 3]{moeini2025-2} and \cite[Sec.~III.C]{8006660}.
We consider the code from Lemma \ref{lem-expurgation-jscc}, and use the following decoding metric
	\begin{align}
		q(\bs{v}, \bs{x}, \bs{y})=W^n(\bs{y} | \bs{x}) \, \exp\big(-2 k D\left(P_i \| P_V\right)\big),
	\end{align}
	if $\bs{v} \in \mcl{T}^k(P_i)$.
	Observe that this is a type-dependent metric, meaning that it depends on $(\bs{v}, \bs{x}, \bs{y})$ only through their types.
	Then, for any $s > 0$, we can upper bound $p_e\big(\bs{v}, j, \msf{C}\big)$ as
\begin{align}
	\!\!\!p_e\big(\bs{v}, j, \msf{C}\big)\!
	&= \mathbb{P}\!\left[\bigcup_{\bar{\bs{v}}\in \mcl{T}^k(P_j)}  q(\bs{\bar{v}}, \bs{x}_{\bar{\bs{v}}}, \bs{Y})  \geq  q(\bs{v}, \bs{x}_{\bs{v}}, \bs{Y}) \right]\\
	&= \mathbb{P}\!\left[\bigcup_{\bar{\bs{v}}\in \mcl{T}^k(P_j)} \!\!  q(\bs{\bar{v}}, \bs{x}_{\bar{\bs{v}}}, \bs{Y})^s  \geq  q(\bs{v}, \bs{x}_{\bs{v}}, \bs{Y})^s \right]\\
	& \leq \sum_{\bs{y}} W^n(\bs{y}|\bs{x}_{\bs{v}}) \!\! \sum_{\bar{\bs{v}}\in \mcl{T}^k(P_j)} \left(\frac{q(\bs{\bar{v}}, \bs{x}_{\bar{\bs{v}}}, \bs{y})}{q(\bs{v}, \bs{x}_{\bs{v}}, \bs{y})} \right)^s.
\end{align}
Substituting the decoding metric and setting $s = \frac{1}{2}$, we obtain
\begin{align}
	\begin{split}
		p_e\big(\bs{v}, j, \msf{C}\big)\,
		&\leq \left(\frac{\exp \left(-k D\left(P_j \| P_V\right)\right)}{\exp \left(-k D\left(P_i \| P_V\right)\right)}\right) \\
		&\hspace{0.05cm}\times \sum_{\bar{\bs{v}} \in \mcl{T}^k(P_j)} \sum_{\bs{y}} \sqrt{W^n(\bs{y} | \bs{x}_{\bs{v}}) W^n(\bs{y} | \bs{x}_{\bar{\bs{v}}})}\,.
	\end{split}
\end{align}
If we assume $\rho_{ij} \geq 1$ for all $(i, j)$, we can use Hölder's inequality \cite[Problem~4.15.f]{gallager} to move the summation over $\bs{\bar{v}}$ outside, as follows 
\begin{align}
	\begin{split}
		\!\!p_e\big(\bs{v}, j, \msf{C}\big)^{\frac{1}{\rho_{ij}}} \!&\leq 
		\vast[\left(\frac{\exp \left(-k D\left(P_j \| P_V\right)\right)}{\exp \left(-k D\left(P_i \| P_V\right)\right)}\right) \\
		&\!\!\!\!\! \sum_{\bar{\bs{v}} \in \mcl{T}^k(P_j)} \! \sum_{\bs{y}} \sqrt{W^n(\bs{y} | \bs{x}_{\bs{v}}) W^n(\bs{y} | \bs{x}_{\bar{\bs{v}}})} \vast]^{\frac{1}{\rho_{i j}}} 
	\end{split}\\
	\begin{split}
		&\leq \left(\frac{\exp (-k D(P_j \| P_V))}{\exp(-k D(P_i \| P_V))}\right)^{\frac{1}{\rho_{i j}}} \\
		&\!\!\!\!\!\sum_{\bar{\bs{v}} \in \mathcal{T}^k(P_j)}\,  \prod_{l=1}^n\left[\sum_y \sqrt{W(y | x_{\bs{v}}^l) W(y | x_{\bs{\bar{v}}}^l)} \right]^{\frac{1}{\rho_{i j}}}\!\!,
	\end{split}
\end{align}
where $x_{\bs{v}}^l$ denotes the $l$-th element of $\bs{x}_{\bs{v}}$.
We now take the expectation of the above expression
\begin{align}
	\begin{split}
		\mathbb{E}\left[p_e\big(\bs{v}, j, \msf{C}\big)^{\frac{1}{\rho_{ij}}}\right] &\leq 
		\left(\frac{\exp (-k D(P_j \| P_V))}{\exp(-k D(P_i \| P_V))}\right)^{\frac{1}{\rho_{i j}}}\\
		&\,\,\,\,\mathbb{E}\left[\sum_{\bar{\bs{v}} \in \mathcal{T}^k(P_j)}  \prod_{l=1}^n \, e^{-\frac{d_B(X_l,\bar{X}_l)}{{\rho_{ij}}}}\right],
	\end{split}
\end{align}
where the expectation is taken with respect to $P_{\bs{X}}^{(\hat{i})} \times P_{\bs{X}}^{(\hat{j})}$,  with $P_{\bs{X}}^{(c)}$ denoting $Q_c^n$.
Here, $\hat{i}$ denotes the index of the class to which source type $i$ belongs.
We can evaluate this expectation as follows
\begin{align}
	\begin{split}
		&\mathbb{E}\left[\sum_{\bar{\bs{v}} \in \mathcal{T}^k(P_j)}  \prod_{l=1}^n \, e^{-\frac{d_B(X_l,\bar{X}_l)}{{\rho_{ij}}}}\right]\\
		&\hspace{1cm}=\sum_{(\bs{x},\bar{\bs{x}})} Q_{\hat{i}}^n(\bs{x}) Q_{\hat{j}}^n(\bs{\bar{x}}) \sum_{\bar{\bs{v}} \in \mathcal{T}^k(P_j)}  \prod_{l=1}^n \, e^{-\frac{d_B(x_l,\bar{x}_l)}{{\rho_{ij}}}}
	\end{split}\\
	&\hspace{1cm}= \sum_{\bar{\bs{v}} \in \mathcal{T}^k(P_j)}  \prod_{l=1}^n \sum_{({x},\bar{{x}})} Q_{\hat{i}}({x}) Q_{\hat{j}}({\bar{x}}) \, e^{-\frac{d_B(x,\bar{x})}{{\rho_{ij}}}}\\
	&\hspace{1cm}\doteq e^{nR_j} \left[\sum_{({x},\bar{{x}})} Q_{\hat{i}}({x}) Q_{\hat{j}}({\bar{x}}) \, e^{-\frac{d_B(x,\bar{x})}{{\rho_{ij}}}}\right]^n.
\end{align}
Raising both to the power $\rho_{ij}$, and then taking the worst-case $Q_{\hat{i}}$ over $\Q_m$ follows that
\begin{align}
	\begin{split}
		\!\!\!\!\!\!\mathbb{E}\left[p_e\big(\bs{v}, j, \msf{C}\big)^{\frac{1}{\rho_{ij}}}\right]^{\rho_{ij}} \!\!\!&\leq 
		\left(\frac{\exp (-k D(P_j \| P_V))}{\exp(-k D(P_i \| P_V))}\right)\\
		&\exp\left(n\!\left[\rho_{ij}R_j - E_{\mathrm{x}}^{\prime}\big(Q_{\hat{j}}, \Q_m, \rho_{ij}\big)\right]\right).
	\end{split}
\end{align}
Having evaluated the expectation, we now obtain the following bound on the error probability of the code in Lemma~\ref{lem-expurgation-jscc} as follows
\begin{align}
	p_e(\mcl{C}) 
	&= \sum_{\bs{v}} P^k(\bs{v})\, p_e(\bs{v},\mcl{C})\\
	\begin{split}
		&\,\, \dot{\leq}\,\, \sum_{(i,j)} \sum_{\bs{v}\in \mcl{T}^k(P_i)}  P^k(\bs{v}) \left(\frac{\exp (-k D(P_j \| P_V))}{\exp(-k D(P_i \| P_V))}\right) \\
	 	&\,\,\times \, \exp\left(-n \left[E_{\mathrm{x}}^{\prime}({Q_{\hat{j}}}, \Q_m, \rho_{ij}) - \rho_{ij}R_j \right]\right). 
	\end{split}
\end{align}
Observe that $ \sum_{\bs{v}\in \mcl{T}^k(P_i)} P^k(\bs{v}) \doteq \exp(-k D(P_i \| P_V))$. This implies that
\begin{align}
	\!\!\!\!\!\!\!\!\!\sum_{\bs{v}\in \mcl{T}^k(P_i)} \!\!\!\!\!\!  P^k(\bs{v})\! \left(\frac{\exp (-k D(P_j \| P_V))}{\exp(-k D(P_i \| P_V))}\right) &\doteq \!\!\!\!\! \sum_{\bs{v}\in \mcl{T}^k(P_j)} \!\!\!  P^k(\bs{v}).
\end{align}
Moreover, since $\prod_{l=1}^k \hat{P}_{\bs{v}}(v_l) = {\exp(-nR_j)}$ for $\bs{v}\in \mcl{T}^k(P_j)$, we obtain
\begin{align}
	p_e(\mcl{C}) 
	&\,\, \dot{\leq}\,\, \sum_{(i,j)=1}^{N_k}\, \sum_{\bs{v}\in \mcl{T}^k(P_j)} \frac{P^k(\bs{v})}{\hat{P}_{\bs{v}}(\bs{v})^{\rho_{ij}}  } \,\, e^{-nE_{\mathrm{x}}'\big({Q_{\hat{j}}}, \Q_m, \rho_{ij}\big)}.
\end{align}
Let us set $\rho_{ij} = \rho_{\hat{j}}$ for all $i$. It then follows that
\begin{align}
	p_e(\mcl{C}) 
	&\,\, \dot{\leq}\,\, \sum_{j=1}^{N_k} \sum_{\bs{v}\in \mcl{T}^k(P_j)} \frac{P^k(\bs{v})}{\hat{P}_{\bs{v}}(\bs{v})^{\rho_{\hat{j}}}  } \,\, e^{-nE_{\mathrm{x}}'\big({Q_{\hat{j}}}, \Q_m, \rho_{\hat{j}}\big)}\\
	& =\,\, \sum_{c=1}^{m}\left[ \sum_{\bs{v}\in \mcl{A}_c} \frac{P^k(\bs{v})}{\hat{P}_{\bs{v}}(\bs{v})^{\rho_{c}}}\right] \,\, e^{-nE_{\mathrm{x}}'\big(Q_{c}, \Q_m, \rho_{c}\big)}.
\end{align}
It was shown in \cite[Corollary~1]{moeini2025-2} that
\begin{align}
	 \sum_{\bs{v} \in \mcl{A}_c}\frac{P^k(\bs{v})}{\hat{P}_{\bs{v}}(\bs{v})^{\rho}} 
	&\,\, \dot{\leq} \,\, \left[\sum_{\bs{v} \in \mcl{A}_c} P^k(\bs{v})^{\frac{1}{1+\rho}}\right]^{1+\rho}.
\end{align}
Thus, we have
\begin{align}
	 p_e \,\, \dot{\leq} \, \sum_{c=1}^m \, e^{-n E_{\mathrm{x}}^{\prime}\big(Q_c, \Q_m, \rho_c\big) +  E_s^{(c)}\big(\rho_c, P^k\big)}.
\end{align}


\subsection{Proof of Theorem \ref{thm-main-1}}
To prove Theorem \ref{thm-main-1}, 
we first state the following auxiliary lemma, given in \cite[Lemma~1]{6803047}.
\begin{lemma}\label{lem-aux}
For any $\rho_0 \geq 1$ and $\gamma^{\prime} \geq 0$, the partition \eqref{eq-part-1}-\eqref{eq-part-2} with $\gamma=\min \left\{1, \gamma^{\prime}\right\}$ satisfies
\begin{align}
	\begin{split}
		&\frac{1}{k} E_{{s}}^{(1)}\left(\rho, P^k\right)\\
		&\hspace{0.5cm} \leq E_{{s}}(\rho, P_V)\,  \mathds{1}\left\{\rho>\rho_0\right\}+r(\rho, \rho_0, \gamma^{\prime}) \mathds{1}\left\{\rho \leq \rho_0\right\}
	\end{split}\\
	&\hspace{0.5cm} \triangleq \bar{E}_{{s}}^{(1)}\left(\rho, \rho_0, \gamma^{\prime}\right), \\
	\begin{split}
		&\frac{1}{k} E_{\mathrm{s}}^{(2)}\left(\rho, P^k\right)\\
		&\hspace{0.5cm} \leq E_{{s}}(\rho, P_V) \mathds{1}\left\{\rho<\rho_0\right\}+r(\rho, \rho_0, \gamma^{\prime}) \mathds{1}\left\{\rho \geq \rho_0\right\} 
	\end{split}\\
	&\hspace{0.5cm} \triangleq \bar{E}_{{s}}^{(2)}\left(\rho, \rho_0, \gamma^{\prime}\right),
\end{align}
where
\begin{align}
	\!\!\!\!\!\!r\left(\rho, \rho_0, \gamma\right) \triangleq E_{\mathrm{s}}(\rho_0, P_V)\!+\!\frac{E_{\mathrm{s}}(\rho_0, P_V)-\log \gamma}{1+\rho_0}\left(\rho-\rho_0\right) .
\end{align}
\end{lemma}
Using this lemma, for any $\gamma’ \geq 0$ the achievable error exponent of Lemma \ref{lem-main} can be evaluated as
\begin{align}
	\begin{split}
		\!\!E_n \!&=\! \liminf_{n\rightarrow \infty} -\frac{1}{n} \log \vast(\! \sum_{c=1,2}\! \exp\!\bigg(-\sup_{\rho_c \geq 1}\Big[n E_{\mathrm{x}}^{\prime}\big(Q_c, \Q, \rho_c\big)\\
		&\hspace{4.15cm}   -  E_s^{(c)}\big(\rho_c, P^k\big)\Big]\bigg) \vast)
	\end{split}\\
	\begin{split}
		&= \liminf_{n\rightarrow \infty}\Bigg\{ \min_{c=1,2} \Bigg\{ \sup_{\rho_c\geq 1}\Bigg\{ E_{\mathrm{x}}^{\prime}\big(Q_c, \Q, \rho_c\big)\\
		& \hspace{3.8cm}- \frac{1}{n}E_s^{(c)}\big(\rho_c, P^k\big)\Bigg\} \Bigg\} \Bigg\}
	\end{split}\\
	\begin{split} \label{eq-apply-lem-aux}
		&\geq \min_{c=1,2} \Bigg\{ \sup_{\rho_0,\rho_1,\rho_2\geq 1}\Bigg\{ E_{\mathrm{x}}^{\prime}\big(Q_c, \Q, \rho_c\big)\\
		& \hspace{3.8cm}- t\bar{E}_{{s}}^{(c)}\big(\rho_c, \rho_0, \gamma^{\prime}\big)\Bigg\} \Bigg\},
	\end{split}
\end{align}
where in \eqref{eq-apply-lem-aux} we applied Lemma \ref{lem-aux} and used the fact that $\liminf _{n \rightarrow \infty} \max _x\left\{f_n(x)\right\} \geq \max _x\left\{\lim _{n \rightarrow \infty} f_n(x)\right\} $.
Optimizing over $\gamma' \geq 0$, swapping the order of $\min$ and $\max$, and restricting the maximization range yields
\begin{align}
	\begin{split}\label{eq-min-term-1}
		\!\!E_n \!&\geq \sup_{\substack{ \rho_0,\rho_1,\rho_2\geq 1\\\rho_1\leq\rho_0\leq\rho_2 }} \sup_{\gamma' \geq 0} \Bigg\{ \min_{c=1,2} \Bigg\{ E_{\mathrm{x}}^{\prime}\big(Q_c, \Q, \rho_c\big)\\
		& \hspace{2.8cm}- t\bar{E}_{{s}}^{(c)}\big(\rho_c, \rho_0, \gamma^{\prime}\big)\Bigg\} \Bigg\}.
	\end{split}
\end{align}
By substituting the expression of $\bar{E}_s^{(c)}(\rho_c, \rho_0, \gamma’)$, the minimization becomes
\begin{align}
	\begin{split}	\label{eq-min-term}
		&\min_{c=1,2} \Bigg\{ E_{\mathrm{x}}^{\prime}\big(Q_c, \Q, \rho_c\big) \\
		&\hspace{0.4cm}+t \frac{E_{{s}}(\rho_0, P_V)-\log \gamma^{\prime}}{1+\rho_0}\left(\rho_0-\rho_c\right)-t E_{{s}}(\rho_0, P_V)\Bigg\}.
	\end{split}
\end{align}
We define $\gamma_0 \geq 0$ as the value satisfying
\begin{align}
	\!\!\!t \frac{E_{{s}}(\rho_0, P_V)-\log \gamma_0}{1+\rho_0}=\frac{E_{\mathrm{x}}'(Q_2,\mcl{Q},\rho_2)-E_{\mathrm{x}}'(Q_1,\mcl{Q},\rho_1)}{\rho_2-\rho_1} .
\end{align}
Choosing $\gamma’ = \gamma_0$ equalizes the two terms in the minimization in \eqref{eq-min-term}, thereby maximizing the lower bound in \eqref{eq-min-term-1}. Consequently, we obtain
\begin{align}
	\begin{split}
		\!\!E_n &\geq \sup_{\substack{ \rho_0,\rho_1,\rho_2\geq 1\\\rho_1\leq\rho_0\leq\rho_2 }} \vast\{  \Bigg\{
		 \frac{\rho_2-\rho_0}{\rho_2-\rho_1}  E_{\mathrm{x}}^{\prime}\big(Q_1, \Q, \rho_1\big)\\
		&\hspace{0.5cm}+\frac{\rho_0-\rho_1}{\rho_2-\rho_1} E_{\mathrm{x}}^{\prime}\big(Q_2, \Q, \rho_2\big)  \Bigg\} - tE_s(\rho_0,P_V)\vast\}
	\end{split}\\
	\begin{split}
		&= \sup_{\rho_0 \geq 1}   \vast\{\sup_{\substack{\rho_1,\rho_2\geq 1\\\rho_1\leq\rho_0\leq\rho_2 }}  \Bigg\{
		 \frac{\rho_2-\rho_0}{\rho_2-\rho_1}  E_{\mathrm{x}}^{\prime}\big(Q_1, \Q, \rho_1\big)\\
		&\hspace{0.5cm}+\frac{\rho_0-\rho_1}{\rho_2-\rho_1} E_{\mathrm{x}}^{\prime}\big(Q_2, \Q, \rho_2\big)  \Bigg\} - tE_s(\rho_0,P_V)\vast\}.
	\end{split}
\end{align}
We can further optimize the expression by assigning the distributions to the classes in the best possible way, obtaining
\begin{align}
	\begin{split}
		\!\!E_n &\geq \sup_{\rho_0 \geq 1}   \vast\{\sup_{\substack{\rho_1,\rho_2\geq 1\\\rho_1\leq\rho_0\leq\rho_2 }}  \Bigg\{
		 \frac{\rho_2-\rho_0}{\rho_2-\rho_1}  E_{\mathrm{x}}^{\prime}\big(\Q, \rho_1\big)\\
		&\hspace{1cm}+\frac{\rho_0-\rho_1}{\rho_2-\rho_1} E_{\mathrm{x}}^{\prime}\big(\Q, \rho_2\big)  \Bigg\} - tE_s(\rho_0,P_V)\vast\}.
	\end{split}
\end{align}
Let $\lambda \in [0,1]$ be defined such that $\lambda \rho_1 + (1-\lambda)\rho_2 = \rho_0$, yielding
\begin{align}
	\begin{split}
		\!\!E_n &\geq \sup_{\rho_0 \geq 1}   \vast\{\sup_{\substack{\rho_1,\rho_2\geq 1, \lambda\in[0,1]\\ \lambda\rho_1+(1-\lambda)\rho_2=\rho_0 }}  \Bigg\{
		\lambda  E_{\mathrm{x}}^{\prime}\big(\Q, \rho_1\big)\\
		&\hspace{1cm}+(1-\lambda) E_{\mathrm{x}}^{\prime}\big(\Q, \rho_2\big)  \Bigg\} - tE_s(\rho_0,P_V)\vast\}
	\end{split}\\
	&=\sup_{\rho_0 \geq 1} \bigg\{ \overline{E}_{\mathrm{x}}'\big(\Q, \rho_0\big) - tE_s\big(\rho_0, P_V\big)\bigg\},
\end{align}
where
\begin{align}
	\!\!\!\!
	\overline{E}_{\mathrm{x}}'\big(\Q, \rho\big) \triangleq  \!\!\!\!\!\!\!
	\sup_{\substack{\rho_1,\rho_2\geq 1, \lambda\in[0,1]\\\lambda\rho_1+(1-\lambda)\rho_2=\rho}} \!\!\!\!\!
	\lambda  E_{\mathrm{x}}^{\prime}\big(\Q, \rho_1\big) + (1-\lambda) E_{\mathrm{x}}^{\prime}\big(\Q, \rho_2\big).
\end{align}
This corresponds to the pointwise supremum over convex combinations of any two values of the function $E_{\mathrm{x}}^{\prime}(\Q, \rho)$, 
which, as described in \cite{rockafellar1970a}, is equivalent to its concave hull. This proof closely follows the argument of \cite[Theorem~2]{6803047}.

\subsection{Proof of Theorem \ref{thm-single-ach}}
The proof follows the same steps as \cite[Corollary~4]{moeini2025-2}, with the main difference being that i.i.d. coding ensembles are used instead of constant-composition.

\subsection{Proof of Theorem \ref{thm-main}}
Let $Q^*$ denote the optimal distribution for $E_{J, \mathrm{ex}}^{\mathrm{sc}}(t)$, namely the solution to
\begin{align}
	E_{J, \mathrm{ex}}^{\mathrm{sc}}(t) = \max_Q \sup _{\rho \geq 1}\bigg\{E_{\mathrm{x}}(Q, \rho)-t E_{{s}}\big(\rho, P_V\big)\bigg\}.
\end{align}
We can lower bound $E_{J, \mathrm{ex}}^{\mathrm{mc}}(t)$ by choosing the maximization pair to be the (potentially suboptimal) choice $\{Q^*,Q^*\}$, as follows
\begin{align}
	E_{J, \mathrm{ex}}^{\mathrm{mc}}(t) 
	&= \max_{\{Q,Q'\}} E_{J, \mathrm{ex}}^{\mathrm{mc}}(\{Q,Q'\}, t)\\
	&\geq  E_{J, \mathrm{ex}}^{\mathrm{mc}}(\{Q^*, Q^*\}, t)\\
	&= \sup_{\rho \geq 1} \bigg\{\overline{E}_{\mathrm{x}}'\big(\{Q^*, Q^*\}, \rho\big) - tE_s\big(\rho, P_V\big) \bigg\}\\
	&= \sup_{\rho \geq 1} \bigg\{E_{\mathrm{x}}'\big(\{Q^*,Q^*\}, \rho\big) - tE_s\big(\rho, P_V\big) \bigg\} \label{eq-conchull-red}\\
	&= \sup_{\rho \geq 1} \bigg\{E_{\mathrm{x}}(Q^*, \rho) - tE_s\big(\rho, P_V\big) \bigg\} \\
	&= E_{J, \mathrm{ex}}^{\mathrm{sc}}(t).
\end{align}
Equality in \eqref{eq-conchull-red} holds because for fixed $Q$, the function $E_{\mathrm{x}}(Q,\rho)$ is concave in $\rho$, so ${E}_{\mathrm{x}}’(\{Q^*,Q^*\},\rho)$ is concave as well, and hence equal to its concave hull.

%

\bibliographystyle{IEEEtran}
\bibliography{refs}

\end{document}